\documentclass[sigconf]{acmart}

\copyrightyear{2025}
\acmYear{2025}
\setcopyright{cc}
\setcctype{by-nc-sa}
\acmConference{Access InContext Workshop @ CHI’25}{April 26}{Yokohama, Japan}
\acmBooktitle{Access InContext Workshop @ CHI’25, April 26, 2025, Yokohama, Japan}

\begin{document}

\title{Refining Participatory Design for AAC Users}

\author{Blade Frisch}
\email{bwfrisch@mtu.edu}
\affiliation{%
  \institution{Michigan Technological University}
  \city{Houghton}
  \state{Michigan}
  \country{USA}
}

\author{Keith Vertanen}
\email{vertanen@mtu.edu}
\affiliation{%
  \institution{Michigan Technological University}
  \city{Houghton}
  \state{Michigan}
  \country{USA}
}

\begin{abstract}
    Augmentative and alternative communication (AAC) is a field of research and practice that works with people who have a communication disability. One form AAC can take is a high-tech tool, such as a software-based communication system. Like all user interfaces, these systems must be designed and it is critical to include AAC users in the design process for their systems. A participatory design approach can include AAC users in the design process, but modifications may be necessary to make these methods more accessible. We present a two-part design process we are investigating for improving the participatory design for high-tech AAC systems. We discuss our plans to refine the accessibility of this process based on participant feedback.
\end{abstract}



\maketitle

\section{Introduction}

Augmentative and alternative communication (AAC) is a field of research and practice that supplements or compensates for a communication disability \cite{american_speech-language-hearing_association_augmentative_nodate, beukelman_augmentative_2020}. AAC comes in multiple forms, such as unaided AAC, where the person uses only their body to communicate, and aided AAC, where an external tool is used. This external tool could be low-tech, such as a picture book or pen and paper, or high-tech, such as a mobile application. Many aided AAC systems, whether low-tech or high-tech, will go through a design process to create the tool.

Blackstone et al. \cite{blackstone_key_2007} proposed six principles for including AAC users in research and practice. These principles state, in part, that it is critical to include people who use AAC in this design process because AAC's social validity is determined by AAC users. However, traditional user research methods might not take into account the challenges that can come with using AAC. For example, AAC users communicate slower than people using natural speech and can get fatigued when using their system, impacting how much and for how long they can communicate. This means methods that take extended amounts of time, such as interviews, may need to be split across multiple sessions to gather the same amount of data as that gathered from people using speech as their primary means of communication. There has been some exploration of including AAC users in general qualitative research, such as making focus groups and interviews asynchronous and including close communication partners to assist with interactions  \cite{beneteau_who_2020}. Still, more work needs to be done on how to include them in design research specifically, such as using co-design to make a new user interface.

\subsection{Participatory Design}

One way to include AAC users in the design process is by taking a participatory design approach. Participatory design is the philosophy that users should be involved in designing the tools they will use by participating in mutual learning and collaboration with designers \cite{muller_participatory_1993, simonsen_routledge_2013}. Originating in a Scandinavian workplace democracy movement in the 1970s, participatory design developed out of a desire to create better tools in an environment becoming more and more shaped by computers \cite{simonsen_routledge_2013}. To do this, users are viewed as sources of knowledge whose lived experiences are vital for shaping the design of technology to be more usable.

There are different ways to implement participatory design that involve different levels of user engagement. An ethnographic approach can be taken to understand technology in its target environment, observing and interacting with users where they use the technology \cite{blomberg_ethnographic_1993}. Contextual inquiry can be used to dig further into what users are doing when performing their tasks, gaining greater awareness and understanding of how the work is being done \cite{holtzblatt_contextual_1993}. Cooperative prototyping, or co-designing, involves the greatest amount of user participation and is where users and designers come together, such as in design workshops, to create the designs for new technology together \cite{bodker_cooperative_1993, rogers_interaction_2023}.

\section{Current Work}

We are using participatory design to guide the creation of a new high-tech AAC system. This AAC system will be designed to better support the social communication and community engagement of autistic adults. Autism is characterized, in part, by a communication disability \cite{american_psychiatric_association_diagnostic_2022}. With approximately 1 in 36 children being diagnosed with autism \cite{maenner_prevalence_2023}, their communication must be supported in all stages of life. We are using a two-step approach in our participatory design implementation. The first step is to run a focus group, a traditional ethnographic research method where multiple participants come together to respond to a series of prompts as a group. The goal of the focus group is to understand how autistic adults engage with their community. Next, we will use a series of design workshops to co-design a new AAC system. However, focus groups and design workshops are not always accessible to AAC users \cite{beneteau_who_2020}. Part of our research is to gather data on the participants' experiences to refine these methodologies to be more accessible in future applications of participatory design methods.

\subsection{Focus Group}
We are currently running the focus group online, asynchronously, and through text communication using Flarum\footnote{\url{https://flarum.org/}}, a free, open-source, and customizable discussion board tool. We chose Flarum over other communication tools, such as Discord or Slack, because it allowed us to preserve participants' anonymity, limit interactions to only the discussion boards, and place restrictions on the ability to edit responses after replies have been received. The participants are autistic adults with various levels of AAC use. We chose this methodology because a focus group allows participants to interact with each other, allowing for common ideas and experiences to emerge across participants. We made the group asynchronous and text-based because AAC users can benefit from asynchronous communication, allowing them more time to compose messages and communication \cite{beneteau_who_2020, kane_at_2017, caron_social_2017}. The current literature has guided these decisions but we are also gathering feedback from the focus group participants on their experience with this methodology. We are collecting data about their ability to share their thoughts and engage with the other participants, what aspects of the methodology they would keep or change, and thoughts on their overall experience taking part in an online, asynchronous, text-based focus group.

\subsection{Design Workshops}
The second step will be to run a series of design workshops. We will recruit autistic adults as well as other AAC stakeholders, such as speech-language pathologists and community support professionals. These workshops will have participants create sketches of new AAC designs. They will do this either individually or with a partner if they are not physically able to sketch themselves. We will collaboratively refine these sketches into a low-fidelity prototype, which representative users will test. We will then collaboratively refine the design again based on the testing results and create a high-fidelity prototype. We will test and refine the design to create a final prototype for a new AAC system.

These design workshops will also be run online. This will allow us to recruit from a wider participant pool and gather a wider range of viewpoints for the design process. Preliminary results from the focus group show that online communities can also be important to autistic adults, so running the workshops in an online setting will allow us to explore a medium that is important to the users. We will also be gathering data on the participant's experiences in the design workshops, asking about their confidence in the design process and generated design, what the best and most difficult parts of their experiences were, and how their attitude towards the design process changed throughout the workshops. These data will allow us to refine the design workshop process to make it more accessible for AAC users in future applications.

\section{Conclusion}

It is critical to include users in the design process when creating technology. This includes the design of assistive technology, such as AAC. Participatory design methodologies can be used to include AAC users, but modifications may be necessary to make these methodologies more accessible. We are using two design methodologies, a focus group and a series of design workshops, to co-design a new AAC system to support the social communication and community engagement of autistic adults. As part of our implementation, we are gathering data from our participants on their experiences in the design process. We will use our findings to help improve the participatory design methodologies to make them more accessible to AAC users.

\section{Acknowledgements}
This material is based upon work supported by the NSF under Grant No.\,IIS-1750193.

\bibliographystyle{ACM-Reference-Format}
\bibliography{references}

\end{document}